\documentclass[10pt,a4pape,twocolumn]{article}
\usepackage{amsmath, amsthm, amssymb,abstract}
\usepackage[utf8]{inputenc}

\usepackage{graphicx}
\usepackage[siunitx]{circuitikz}
\usepackage{blindtext}
\usepackage{array}
\usepackage{placeins}
\usepackage{subcaption}
\usepackage{pgfplots}
\usepackage{physics}
\usepackage{amsfonts,latexsym,graphicx,enumerate,setspace,verbatim,tocloft,rotating}
\graphicspath{{images/}}
\usepackage[siunitx]{circuitikz}
\usepackage{authblk}
\usepackage[perpage]{footmisc}

\begin{document} 

\title{Initial Design of a W-band Superconducting Kinetic Inductance Qubit}

\author[1,2]{Farzad B. Faramarzi \thanks{ffarama1@asu.edu}}
\author[3]{Peter K. Day\thanks{peter.k.day@jpl.nasa.gov}}
\author[1,5]{ Jacob Glasby \thanks{jglasby@asu.edu}}
\author[2]{ Sasha Sypkens \thanks{ssypkens@asu.edu}}
\author[7]{ Marco Colangelo \thanks{colang@mit.edu}}
\author[1]{ Ralph Chamberlin \thanks{ralph.chamberlin@asu.edu}}
\author[6]{ Mohammad Mirhosseini \thanks{mohmir@caltech.edu}}
\author[1]{ Kevin Schmidt \thanks{kevin.schmidt@asu.edu}}
\author[7]{ Karl K. Berggren \thanks{berggren@mit.edu}}
\author[1,2]{ Philip Mauskopf \thanks{Philip.Mauskopf@asu.edu}}
\affil[1]{Department of Physics, Arizona State University, USA}
\affil[2]{School of Earth and Space Exploration, Arizona State University, USA}
\affil[3]{Jet Propulsion Laboratory, NASA, USA}
\affil[5]{School of Electrical, Computer, and Energy Engineering, Arizona State University, USA}
\affil[6]{Department of Electrical Engineering,
California Institute of Technology, USA }
\affil[7]{Department of Electrical Engineering and Computer Science, Massachusetts Institute of Technology,USA}

\maketitle

\begin{abstract}
Superconducting qubits are widely used in quantum computing research and industry. We describe a superconducting kinetic inductance qubit (and introduce the term Kineticon to describe it) operating at W-band frequencies with a nonlinear nanowire section that provides the anharmonicity required for two distinct quantum energy states. Operating the qubits at higher frequencies may relax the dilution refrigerator temperature requirements for these devices and paves the path for multiplexing a large number of qubits.  Millimeter-wave operation requires superconductors with relatively high $T_c$, which implies high gap frequency, 2$\Delta/h$, beyond which photons break Cooper pairs. For example, NbTiN with $T_c =15\,\text{K}$ has a gap frequency near 1.4 THz, which is much higher than that of aluminum (90 GHz), allowing for operation throughout the millimeter-wave band. Here we describe a design and simulation of a W-band Kineticon qubit embedded in a 3-D cavity. We perform classical electromagnetic calculations of the resulting field distributions.
\end{abstract}

Superconducting qubits are one of the leading platforms for building quantum computers. Recently, quantum supremacy, in which a quantum computer performs a computation which is infeasible on a classical computer, was demonstrated using 53 superconducting qubits \cite{arute_quantum_2019}. Several companies and academic institutions (IBM, Rigetti, QuTech, and Amazon) offer cloud quantum computing services based on superconducting transmon qubits. These devices, operating in the 4-10 GHz band utilize Josephson junctions formed by an aluminum oxide layer between aluminum contacts. Scaling up these early quantum computers to the thousands or millions of physical qubits needed to realize quantum error correction \cite{nas_quantum_2018} faces several major hurdles including limited coherence (10-100 $T_2^*$ is typical), lack of room temperature interconnects, and the large physical size of superconducting qubits (typical qubits have 0.1-1 mm lateral dimensions). The standard aluminum or niobium Josephson tunnel junction is becoming a bottleneck to increasing qubit coherence and yield requirements: Spurious two level systems (TLS) cause decoherence \cite{klimov_fluctuations_2018, burnett_decoherence_2019, schlor_correlating_2019}, imprecision in qubit frequencies reduces yield for fixed frequency qubit architectures \cite{chamberland_topological_2020}, and quasipartlces cause charge parity fluctuations and heating \cite{serniak_hot_2018}. Superconducting kinetic inductance qubits operating in the W-band (75-110 GHz) have the potential to lift these bottlenecks. The nonlinear inductance of superconducting nanowires has been used in designs and experiments involving cavity-based parametric amplifiers \cite{PhysRevLett.103.087003, doi:10.1063/1.3570693} and qubits \cite{winkel_implementation_2019, schon_rabi_2020,Kerman} in the sub-10 GHz regime and as a parametric amplifier in the W-band \cite{anferov}, but thus far, a viable superconducting qubit has not been demonstrated in the W-band.

In particular, exploration of high-frequency qubits is encouraged because they have the potential to solve current problems with state-of-the-art transmon qubits. In order to achieve the highest coherence times, IBM uses fixed-frequency transmons and an all-microwave operation to entangle pairs of qubits together (the cross resonance gate). However, the speed and performance of the cross resonance gate depends heavily on the exact frequency allocation between the qubits \cite{chamberland_topological_2020} and spectator qubits \cite{takita_demonstration_2016}. Fabricating the nonlinear inductive part of the qubit from well-defined lithographic processes of non-linear kinetic superconductors instead of amorphous growth of an aluminum oxide layer in the Josephson junction could allow repeatable frequency allocation between qubits. Also operating at a higher frequency offers the ability to scale the systems down in size as we look to scale to ever larger numbers of qubits. 

Construction of a quantum computer implementing error correction imposes constraints on size, operating frequency, and operating temperature. The computer will require at least thousands of physical qubits, a scale at which the physical size of the circuits residing at milliKelvin temperatures becomes a limitation.   This is particularly true for qubits coupled to 3D microwave cavities, but also holds for 2D circuits.  An interesting possibility is to scale the circuits to a much higher frequency. Millimeter-wave operation requires superconductors with relatively high $T_c$,  which implies high gap frequency, $2\Delta/h$, beyond which photons break Cooper pairs.  For example NbTiN with $T_c \approx 15\,$K \cite{NbTiNTc} has a gap frequency near 1.4~THz, much higher than that of aluminum (90 GHz), allowing for operation throughout the millimeter-wave band. We introduce the new term Kineticon to refer to qubits that take advantage of the nonlinear response of superconducting wires as opposed to relying on Josephson junctions.\\

In this section, we quantize the Kineticon system following closely the standard treatment for quantizing electrical circuits\cite{Devoret,yurk}.

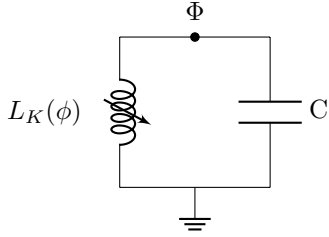
\begin{figure}[tp]
    \centering
    \begin{circuitikz}
      \draw
      (0,-1) node[circ]{}
      (0,-0.65) node{$\Phi$}
      (0,-1)--(-1,-1)
      (-1,-1) to [vL](-1,-3)
      (-2,-2) node{$L_K(\phi)$}
      (0,-1)--(1,-1)
      (1,-1) to [C=C](1,-3)
      (1,-3)--(-1,-3)
      (0,-3) node[ground]{};
    \end{circuitikz}
    \caption{Circuit diagram of an ideal superconducting LC resonator with nonlinear inductive element. }
    \label{circ}
\end{figure}

A Kineticon qubit is very similar to an LC resonator circuit, save for the inductive part of the circuit, which is non-linear. A simple Kineticon qubit circuit can be imagined as in Fig \ref{circ}, with one active node and no current bias. Defining a branch flux at this node, the kinetic inductance of the nanowire in the weak anharmonic limit, $\Phi \approx L_{0k} I$ can be written as \cite{PhilMKIDs} \\
\begin{equation}
L_k(\Phi) \approx L_{0k} \Bigg( 1 + \frac{\Phi^2}{\Phi_*^2} \Bigg)    
\end{equation}\\
where $L_{0k}$ is the kinetic inductance with zero bias and $\Phi_* \equiv I_* L_{0k}$ where $I_*$ is a characteristic current parameter of the nonlinearity.\\

The energy stored in the capacitor and the nonlinear inductor can be found using \\
\begin{equation}
E(t) = \int_{-\infty}^t v_b(t') i_b(t') dt'   
\end{equation}
where $v_b$ and $i_b$ are the voltage and current of the branch, respectively\cite{Devoret}. Calculated energies are as follows\\
\begin{equation}
U_C =\frac{1}{2}C \Dot{\Phi}^2    
\end{equation}
\begin{equation}
U_L \approx E_k \Bigg( \frac{\Phi^2}{\Phi_*^2} - \frac{\Phi^4}{\Phi_*^4} \Bigg)    
\end{equation}
where we define $E_k = \frac{\Phi_*^2}{2 L_{0k}}$. The Lagrangian is given by \\
\begin{equation}
\mathcal{L} = U_C -U_L = \frac{1}{2} C \Dot{\Phi}^2 -  E_k \Bigg( \frac{\Phi^2}{\Phi_*^2} - \frac{\Phi^4}{\Phi_*^4}\Bigg).   
\end{equation}
Now using Legendre transformation $\mathcal{H} = \Dot{\Phi}Q - \mathcal{L}$ we can write the Hamiltonian as follows \\
\begin{equation}
\mathcal{H} = \frac{Q^2}{2C} + E_k  \Bigg( \frac{\Phi^2}{\Phi_*^2} - \frac{\Phi^4}{\Phi_*^4}\Bigg).   
\end{equation}
 We can quantize the circuit by replacing $Q$ and $\Phi$ with their quantum operators that satisfy the following commutation relation \\
 \begin{equation}
[Q,\Phi] = -i \hbar.    
 \end{equation}

\FloatBarrier
The Hamiltonian operator now can be written as follows by replacing charge and flux with their quantum operators
\begin{equation}
\tilde{H} =  \frac{\Tilde{Q}^2}{2C} + \frac{\Tilde{\Phi}^2}{2L_{0k}} - \frac{1}{2 L_{0k} \Phi_*^2} \Tilde{\Phi}^4. 
\end{equation}

In the second quantization language, we can define the following creation and annihilation operators  \\

\begin{equation}
\tilde{a} = \frac{1}{\sqrt{\hbar \omega_r}} \Bigg[ \frac{1}{\sqrt{2 L_{0k}}}\tilde{\Phi} + i  \frac{1}{\sqrt{2 C}}\tilde{Q}   \Bigg]
\end{equation}

\begin{equation}
\tilde{a}^{\dagger} = \frac{1}{\sqrt{\hbar \omega_r}} \Bigg[ \frac{1}{\sqrt{2 L_{0k}}}\tilde{\Phi} - i  \frac{1}{\sqrt{2 C}}\tilde{Q}   \Bigg].
\end{equation}

The reduced charge and flux operators in terms of $\tilde{a}$ and $\tilde{a}^{\dagger}$ are given by \\
\begin{equation}
\tilde{q} = \frac{\tilde{Q}}{ q_{zpf}} =i  \bigg(\tilde{a} - \tilde{a}^{\dagger} \bigg)    
\end{equation}
\begin{equation}
\tilde{\phi} = \frac{\tilde{\Phi}}{\varphi_{zpf}} = \bigg(\tilde{a} + \tilde{a}^{\dagger} \bigg).    
\end{equation}
where $Z_0 = \sqrt{\frac{L_{0k}}{C}}$ and we defined $\varphi_{zpf}=\sqrt{ \frac{\hbar Z_0}{2 }}$ and  $q_{zpf}=\sqrt{\frac{\hbar}{2 Z_0}}$ as zero-point fluctuations of flux and charge respectively. \\

The Hamiltonian of our weakly anharmonic oscillator (AHO) becomes 
\begin{equation}
\tilde{H}= \hbar \omega_r \bigg[ a^{\dagger}a +\frac{1}{2}  + \frac{1}{4} \lambda \bigg(a^{\dagger}+a\bigg)^4\bigg]    
\end{equation}
where $\omega_r= \sqrt{\frac{1}{L_{0k}C}}$ and $\lambda = -\frac{\varphi_{zpf}^2}{\Phi_*^2}$.\\

Addressing individual states in a qubit requires a large relative anharmonicity, $\alpha = |E_{10} - E_{21}| / E_{10}$, where $E_{10}$ is the transition energy from the ground state to the first-excited state and $E_{21}$ is the transition energy from first-excited state to the second-excited state.  Here, we derive this expression for the Kineticon qubit, and discuss the implications of this requirement.

\begin{figure*}[htbp]
    \centering
    \includegraphics[width=0.55\linewidth,height=0.4\linewidth]{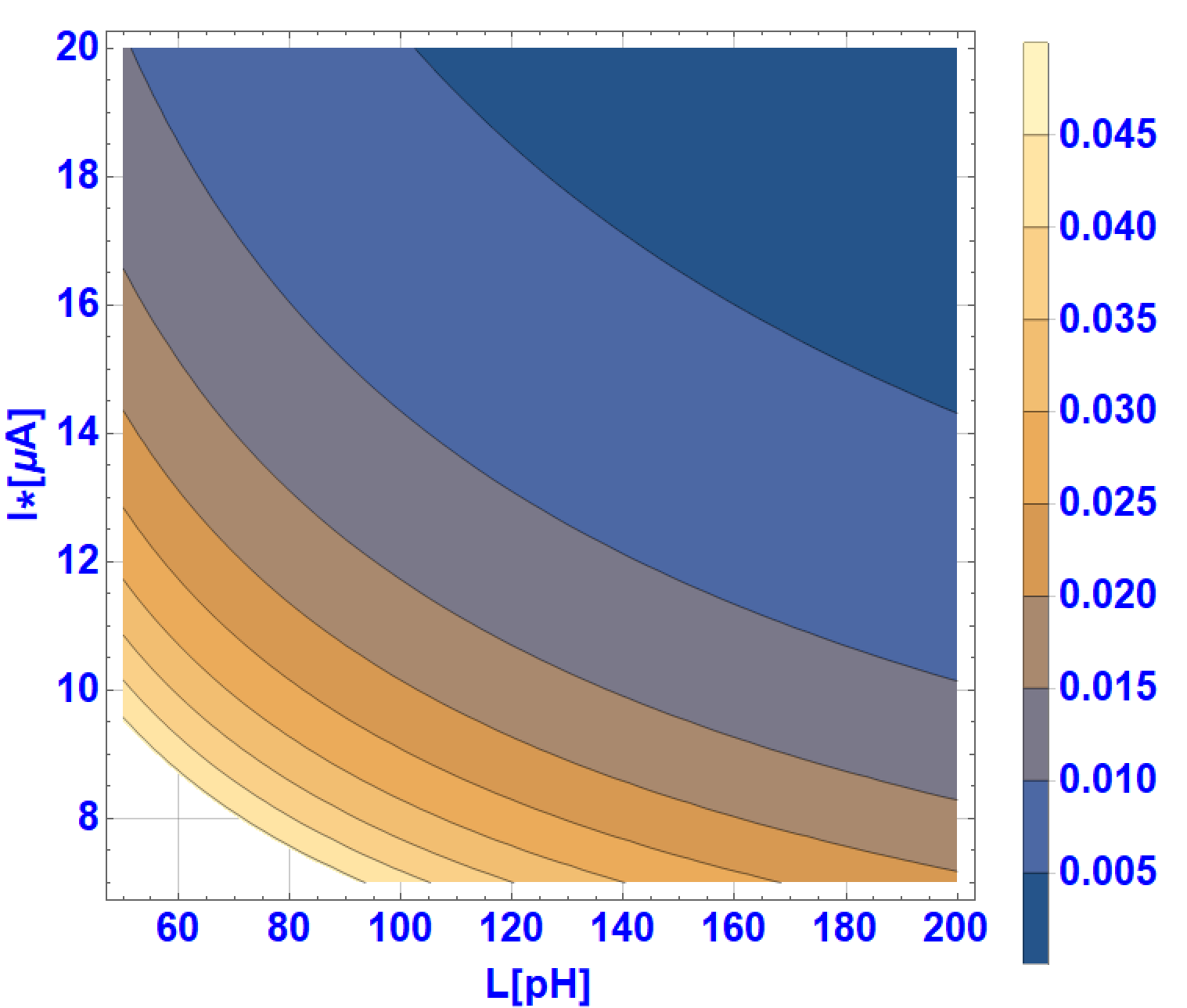}
    \includegraphics[width=0.4\linewidth]{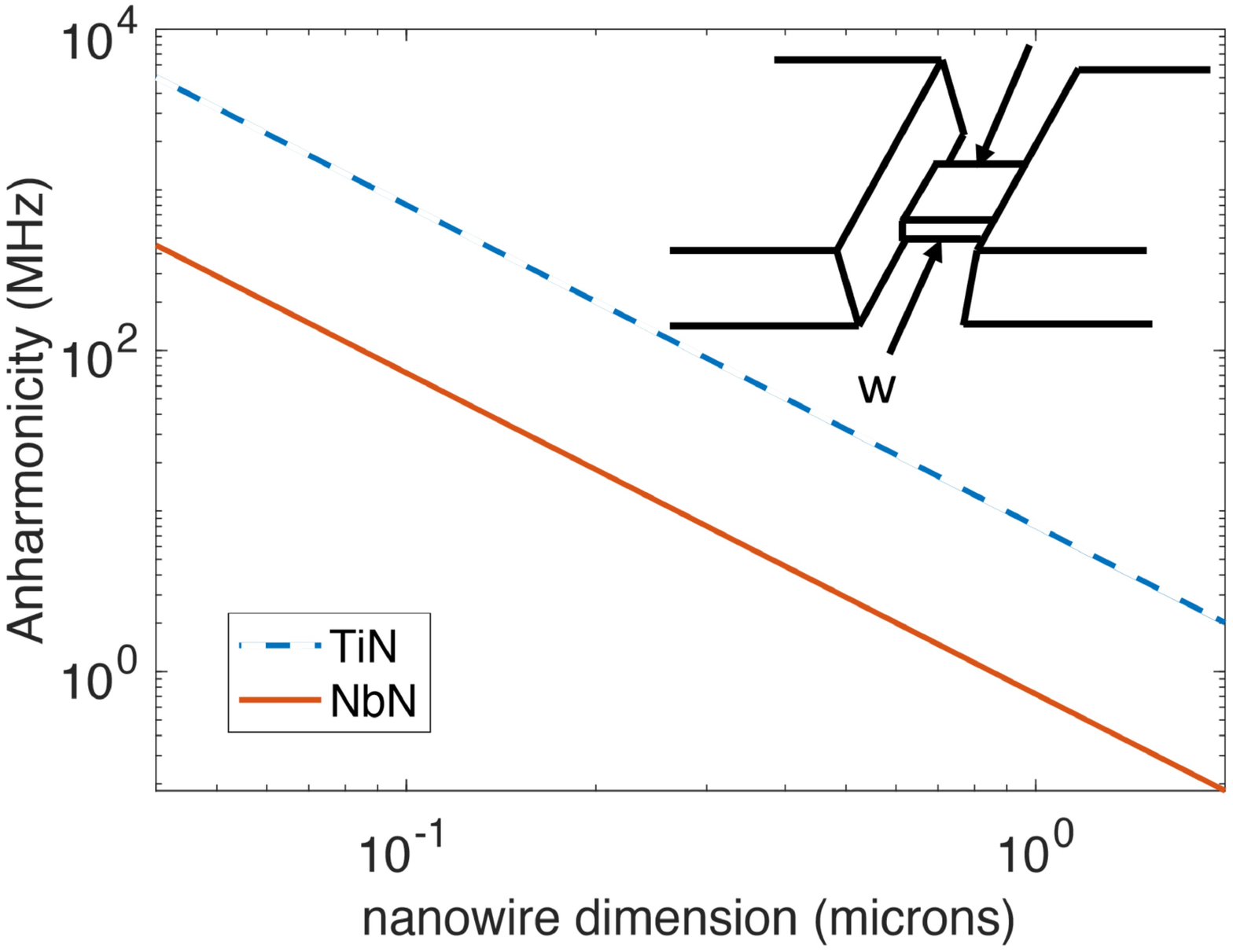}
    \caption{{\em Left:} Contour plot of the relative anharmonicity of a nanowire qubit as a function of total inductance $L$ and the characteristic current $I_*$. {\em Right:}  Anharmonicity versus nanowire dimension for a 100~GHz resonator with an inductance dominated by an embedded TiN and NbN nanowire. The anharmonicity increases with decreasing nanowire dimensions, and is larger for TiN than NbN.}
    \label{fig:nanobridge}
\end{figure*}

Increasing both the qubit and readout resonator frequency to W-band ($\sim 100$~GHz) could allow for operation of the quantum processor at a higher temperature.  While the 90~GHz gap frequency of aluminum represents a barrier for conventional qubit technologies, it may be possible to realize a qubit with a transition frequency in the millimeter band that uses nonlinear kinetic inductance to provide the required anharmonicity. For a Kineticon qubit the relative anharmonicity, $\alpha$ can be written as follows \\ 
\begin{equation}
\alpha \approx 3 \frac{I^2_{zpf}}{I^2_*}    
\end{equation}
where $I_{zpf} = \sqrt{\frac{h f_r}{2 L}}$ is the zero-point fluctuation current and $I_*$ is the characteristic current of the nanowire. The factor 3 comes from energy eigenvalue calculations for a nanowire in the weak anharmonic limit. For a Kineticon qubit with $f_r = 100 $ GHz, we can plot the relative anharmonicity as functions of total inductance of the nanowire $L$ and the characteristic current $I_*$.  As we can see from Fig.\ref{fig:nanobridge}, lowering both total inductance $L$ and the characteristic current $I_*$ increases the relative anharmonicity. To improve the anharmonicity of a Kineticon qubit, it is possible to control $J_*$ and $L_s$ of the nanowire during the fabrication process \cite{Aref_2011,Annunziata_2010} in addition to changing its dimensions.

We can express $I_*$ of the nanowire in terms of the material parameters using Mattis-Bardeen Theory\cite{MBT} and following \cite{zmuidzinas2012} as 
\begin{equation}
    I_* = J_* w t = \sqrt{\frac{\pi N(0) \Delta^3}{\hbar \rho_n}}wt
\end{equation}
where $w$ and $t$ are the width and thickness of the nanowire respectively and $\rho_n$ is the normal state resistivity of the thin film. Using $L_s =\hbar R_s / \pi \Delta $, we can express the relative anharmonicity in terms of the volume, $V$, of the nanowire:
\begin{equation}
\alpha \approx 3 \frac{hf_{r}}{2N(0)\Delta^2 V},
\end{equation}
where $f_r$ is the resonator frequency, $N(0)$ is the density of states at the Fermi level, and $\Delta$ is the gap parameter.  Here we assume the resonator inductance is dominated by that of the nanowire.  The denominator in this expression is the superconducting condensation energy, which may also be expressed in terms of the critical field.  Using $N(0) = 8.7
\times 10^9$eV$^{-1}${\textmu}m$^{-3}$\cite{leduc2010titanium} and $\Delta = 0.5$meV for a thin TiN film ($T_c = 4.5$ K and gap frequency of $f_{gap}\approx$ 265 GHz), $N(0) = 2
\times 10^{10}$ eV $^{-1}${\textmu}m$^{-3}$ and $\Delta = 1.1$ meV \cite{schroeder2018development,semenov2009optical}  for a NbN thin film, and setting a nanowire thickness to 5~nm and the length and width to be equal, the anharmonicity is shown in Fig.~\ref{fig:nanobridge}. 

While an absolute anharmonicity comparable to a transmon ($\sim 200$~MHz) may be achieved with dimensions that are more or less straightforward to produce, the question is whether TLS or other loss mechanisms will contribute a loss tangent of ($\sim 10^{-6}$) for this type of qubit, as they do for current state-of-the-art microwave qubits.  In that case, to maintain the ratio of decay time to read time the anharmonicity would need to be increased accordingly, requiring increased length or decreased width, so the fabrication becomes more challenging. We need to investigate non-linearity in resonators with embedded nanowires in order to better understand the design requirements.\\

Apart from the possibility of high frequency operation, an advantage of circuits made from relatively high T$_c$ materials is that they can be less affected by quasiparticles:
\begin{itemize}
\item The thermal quasiparticle density goes as $e^{-2\Delta/T}$.
\item The mean number of non-thermal quasiparticles produced by absorption of a phonon of a given energy is $\xi h\nu / \Delta$, where $\nu$ is the phonon frequency and $\xi$ is a material dependent parameter (around 0.5 for most materials).
\item Quasiparticles that are created through leakage of radiation into the cryogenic environment recombine at a rate that scales as $T_c^{-3}$ \cite{kaplan1976quasiparticle}.
\end{itemize}
In addition to lower quasiparticle loss, the nitride superconductors that we will study may be affected less by loss and decoherence associated with two level systems.  It is known, for example, that TiN and NbTiN have high quality surfaces and form more stable and thinner oxide layers \cite{oxide}, unlike elemental superconductors such as Al, Nb and Ta.  TiN in particular is a hard material and is used as a coating on machine tools, drill bits, etc.  Groups working on kinetic inductance detectors have found that fabricating the non-photoresponsive parts of the detectors out of NbTiN leads to lower TLS noise \cite{nbtintls1,nbtintls2}.  Recent results on transmon qubits using Ta electrodes have shown an improvement over state-of-the-art coherence times \cite{place2020new}, which was associated with the favorable properties of the tantalum oxide surface, so it is interesting to ask whether the metal nitride superconductors, and other potential materials for Kineticons, will provide further improvement.\\

In this section, we present designs corresponding to two stages of development of the Kineticon qubit.  The first design is based on thin film technology and is most appropriate for early-stage investigations while the second design is a more complete three-dimensional design in which the qubit is coupled to a readout cavity and is more appropriate for later-stage work.\\

\begin{figure*}[htbp]
    \centering
    \includegraphics[width=5 in,height=3 in]{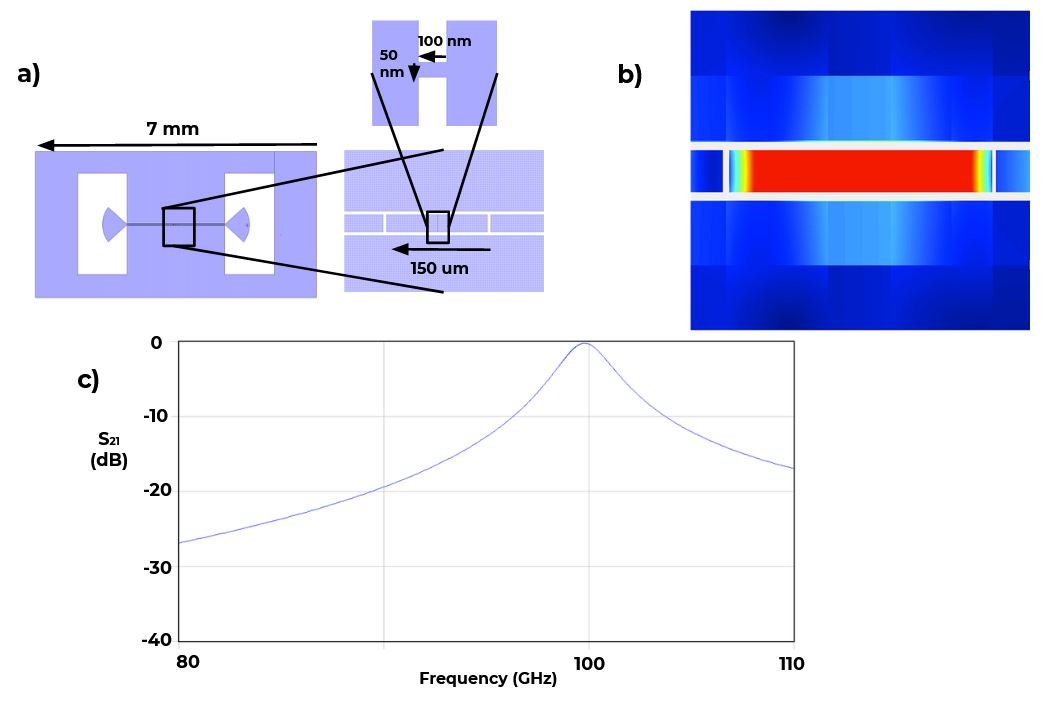}
    \caption{{\em a) A mask layout of a possible design for a nanowire embedded in a 2D Fabry-Perot cavity. The layout includes coupling antennas and coplanar waveguide feed lines}. {\em b) Map of the magnitude of the current density on the resonator at its fundamental frequency simulated using a commercial electromagnetic simulator, Sonnet software \cite{sonnet}. }  {\em c) Simulated $S_{21}$ response of the resonator as a function of frequency.} }
    \label{fig:2d cav}
\end{figure*}

To investigate the anharmonicity in different materials and structures, we have designed a nanowire with different dimensions embedded in a thin-film Fabry-Perot resonator. The nanowire is placed in the middle of the resonator where the current has an anti-node at the fundamental frequency $f_0$ ($\lambda/2$) as shown in Fig.\ref{fig:2d cav}.
Since the inductance of the resonator is dominated by the kinetic inductance of the nanowire, any change in the kinetic inductance caused by adjusting the readout power shifts the resonant frequency. Putting the nanowire in the above mentioned configuration helps us to measure the frequency shift due to very small number of photons in the resonator. For such a qubit we require the frequency shift $\delta f$ to be larger than number of photons $n$ times the bandwidth $B$, $\delta f > n B/2\pi$ \cite{Ku}. \\

To put a W-band tone into the resonator we designed a single mode rectangular waveguide to CPW transition and capacitively couple the signal to the resonator. We use am commercially available electromagnetic simulator Ansys Electronics Suite (HFSS) \cite{hfss} to simulate and optimize the coupling efficiency of the transition\cite{Faramarzi2020}.We expect a coupling efficiency of 70 percent or better at 100 GHz.\\

Eventually, this technology will require coupling to other qubits, which can be achieved by embedding the qubit in a three-dimensional cavity.  That cavity also could reduce the influence of defects in the thin-film system by increasing the mode volume and reducing the field density \cite{3dtransmon}.

\begin{figure*}[htbp]
    \centering
    \includegraphics[width=6 in,height=4 in]{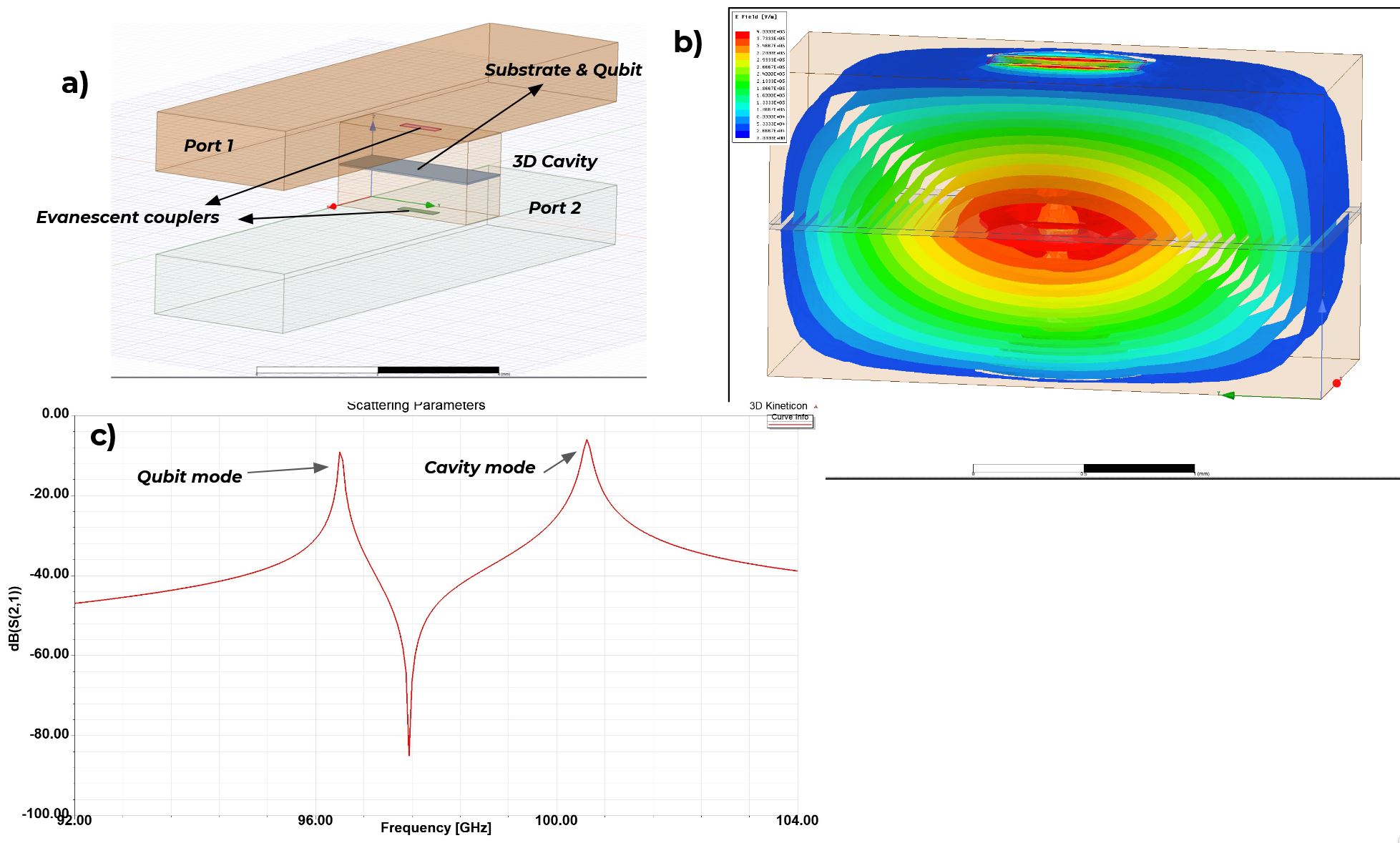}
    \caption{{\em a) HFSS overview drawing of the 3D system of qubit coupled to 3D readout cavity, coupled evanescently to readout waveguides. Gray slab in the middle of the 3D cavity would hold the qubit device.} {\em b) The magnitude of electric at the cavity mode in the 3D readout cavity.  Qubit is located at the center of this cavity, where the electric field is maximum.{\em c) $S_{21}$ calculated }response of the qubit-cavity system showing separation of the qubit and cavity resonances.}  }
    \label{fig:3dcav}
\end{figure*}

To embed the Kineticon qubit in a 3D resonant readout cavity, we have designed and simulated a 100 GHz resonant cavity coupled to two waveguides through evanescent couplers as shown in Fig.\ref{fig:3dcav}.a. The dimensions of the cavity was adjusted such that the resonant frequency of the cavity-substrate system is nearly 100 GHz. In this case the dimensions of the cavity are; width = 1 mm, height = 2.54 and depth = 1.4 mm. The Kineticon qubit is placed in the middle of the cavity where the dipole moment of the qubit is aligned with the $TE_{101}$ mode and where the electric field is maximum. The transmission response of the cavity-qubit system is shown in Fig.\ref{fig:3dcav}.c where both qubit and cavity modes are visible.\\

In conclusion, we have investigated and simulated the possibility of a W-band qubit resonator by taking advantage of the non-linear kinetic inductance of a nanowire as the anharmonic element. Increasing the operating frequency and readout frequency of the qubits relaxes the very low temperature requirement. Higher frequency also means smaller cavity and qubit dimensions, which helps with scaling up the number of qubits for quantum computation.\\

The approach we suggest has a number of possible problems with it: (1) the new materials we plan to use might prove to induce new sources of noise and decoherence; (2) the high frequencies we plan to work at may be challenging to synthesize and control in performing qubit operations; and (3) the operation at higher temperatures and frequencies may introduce different kinds of noise and decoherence that have not been previously observed.  In addition, our calculations so far have been purely classical---a quantum mechanical treatment is clearly in order. To resolve these questions, additional work will be required.

The first step is to fabricate the Fabry-Perot resonator with the embedded nanowire to measure frequency shifts and estimate the anharmonicity for different materials, such as NbN and TiN, and compare them with the analytical expressions derived in this work.  In parallel we plan to fabricate the 3D cavities mentioned here using micro-machining and test them warm and cold. Next step will be to characterize the cavity- qubit system at W-band frequencies.   
\section*{Acknowledgement}
M.C. and K.K.B. acknowledge the support of the National Science Foundation under contract No. ECCS-2000743 (MIT).

\bibliographystyle{unsrt} 
\bibliography{references.bib}

\clearpage

\end{document}